\documentclass{article}
\usepackage{arxiv}
\usepackage[utf8]{inputenc} 
\usepackage[T1]{fontenc}    
\usepackage{hyperref}       
\usepackage{url}            
\usepackage{booktabs}       
\usepackage{amsfonts}       
\usepackage{nicefrac}       
\usepackage{microtype}      
\usepackage{lipsum}		
\usepackage{graphicx}
\usepackage{natbib}
\usepackage{doi}

\title{Short Review: Pathology of the image big data era using electron microscopy}

\author{
    {Satoshi Kume} \\
    Osaka, Japan \\
	\texttt{satoshi.kume@a.riken.jp}
}

\renewcommand{\shorttitle}{\textit{arXiv}}

\begin{document}
\maketitle

\section{Introduction -Fusion of Innovative Imaging Technology and Pathology-}

The human body is mysterious, consisting of innumerable fine and complex structures and a vast amount of information. The field of modern pathology has been systematized around the findings and observations obtained from light microscopy. However, such observations in the visible light range are limited to those at the cellular level; thus, observations of cellular ultrastructures are difficult. Therefore, only vague information can be collected.

Various imaging methodologies and tissue staining techniques have been created to further understand the complexity of biological tissues. Using such approaches, it has been realized that cellular morphology data that comprehensively acquire biological tissue information can be treated as new omics information, termed the 'Totalities of body tissue morphology (morphome)'.

In addition, it is also possible to revisit vague morphological information quantitatively and objectively using knowledge relating to the latest scientific information. Currently, new technologies are allowing additional frontiers to be opened. Indeed, I believe that a new revolution is about to take place regarding the 'viewing' of tissues by innovative imaging technologies such as 'wide-area electron microscopy (EM)', which can observe a wide area of several millimeters or more with nano-level resolution. Moreover, with the advent of cutting-edge information technology, I expect big data processing to be more innovative.

Through the incorporation of these new methodologies, the field of pathology is expected to deepen through the identification of previously unknown structures, the quantification of infrequent rare events, the new reclassification of diseases, and the automatic diagnosis of diseases. In this review, I introduce the innovation of the EM imaging and discuss the field of pathology with regard to the big data era.

\section{Wide-area electron microscopy (EM) to observe biological tissue in a wider area}

\subsection{Conventional EM analysis}

In the electron microscopic observation of biological tissues, resin-embedded section observation methods by transmission electron microscope (TEM) are typically used. As part of this observation method, pre-fixation of the tissue samples with aldehydes are first performed, followed by post-fixation with a heavy metal oxidizing agent such as osmium tetraoxide. Next, dehydration of the biological samples is performed using an alcohol solvent, and the sample is substituted and embedded in an epoxy resin. Then, the resin-embedded tissue sample is ultra-thinned to a thickness of 100 nm or less using an ultramicrotome. The resulting ultra-thin tissue sections are then placed on a metal TEM grid having a 100–150 $\mu$m/pitch mesh structure. Subsequently, electron staining is performed using heavy metals such as uranium acetate or lead. Through the utilization of this staining for electron microscopic observation, intracellular morphology, including the cell nucleus, mitochondria, and Golgi body, along with the nuclear and cell membranes can be observed in clear contrast.

In TEM, an electron beam with an acceleration voltage of approximately 100 kV is applied to a sample, and transmitted or forward scattered electrons are detected. Then, the morphological image of the biological tissue is obtained similar to a shadow picture. By increasing the accelerating voltage that is applied to the electron gun and shortening the wavelength of the radiation source extremely, high resolution at the nano-level becomes possible. For example, when the acceleration voltage is 100 kV, the wavelength of the electron beam is approximately 0.0037 nm. By using this extremely short wavelength region, biological tissue observations at a super resolution of at least 100,000 times magnification are realized. This would be impossible in the visible light region. However, when observing thick tissue sections, TEM images become unclear because of the overlapping of depth information and increased inelastic scattering.

For this reason, during conventional TEM observations, an ultra-thin section with a thickness of 100 nm or less is placed and observed on a TEM grid having a pitch mesh structure. However, currently, one disadvantage is that the ultra-thin tissue section is easily broken on the grid, and the observation area is limited because only the gaps in the grid can be observed. Because of these limitations, conventional electron microscopic analysis has primarily been confined to a narrow area of observation and is considered only as an auxiliary diagnosis in the field of pathology.

\subsection{Next-generation scanning electron microscope (SEM)}

A SEM can detect various scattered components on the incident side by scanning an electron beam on the sample to obtain a high-resolution image. Therefore, SEM represents an electron microscope technique that differs from TEM. As a representative example of the biological observations that can be made using SEM, observations of the surface three-dimensional structure of both bacteria and organs can be acquired. During this process, the electron beam is scanned on a bulk sample surface-coated with osmic acid in order to detect the secondary electron signal. Using this technique, the resulting SEM image will include surface asperity information. However, thus far, the application of SEM to medical fields is limited.

In recent years, trends in ultrastructural analysis have significantly changed. Among the changes, high-resolution SEM equipped with a field emission electron gun has been developed \cite{SUGA2014}. Using this device, it is possible to control the electron probe diameter extremely thinly while maintaining a voltage of low (10 kV or less) or ultra-low acceleration (2 kV or less). As a result, it is possible to observe minute morphological information on the surface of the sample while reducing beam damages to the biological tissue.

Moreover, during SEM observations, since the tissue sample is placed on a silicon wafer substrate, it is possible to observe ultra-thin tissue sections for a longer period of time. In addition, a plurality of detectors can be mounted in the SEM apparatus. Furthermore, in addition to the conventional secondary electron detector, a backscattered electron detector and an X-ray detector can also be used. In particular, due to the dramatically high sensitivity of the backscattered electron detector, high-resolution observations of the ultra-thin slices conventionally observed with TEM are possible even with the use of SEM. By applying this next-generation SEM technology, 'wide-area SEM' and 'stereoscopic SEM' can be realized.

\subsection{Large-scale 2D-SEM}

By using the latest SEM and improving the control system, Kume et al. previously introduced a technology that can image a wide area of approximately 1 mm square using ultra-thin sections of biological tissue with an electron microscopic resolution \cite{Kume2017}. In the wide-area EM, the software control of the sample stage is performed with high precision and automatic imaging is repeated using the autofocus function. For example, tiled image data (tiling image) are acquired by continuous imaging in a zigzag form, whereas wide-area tissue images can be observed by computationally connecting the respective images.

Example imaging data of wide-area SEM using the rat liver are shown (Data not shown). In this image, 5431 image data points were acquired and the whole rat liver leaflet (approximately 0.7 mm × 1.0 mm) was reconstructed. During this observation, the working distance between the sample and the detector was developed under low acceleration voltage conditions (7 to 8 kV), and the reflected electron signal emitted was detected. In addition, under ultra-low acceleration conditions (about 1 to 2 kV), only the signal on the surface of the sample can be obtained. Therefore, even if it is not an ultra-thin slice, even a thick tissue slice (1 $\mu$m or more) can be observed by SEM, and a high-resolution tissue image independent of slice thickness can be obtained. The reverse contrast of the original reflection electron image obtained by SEM sufficiently reproduces the tissue image obtained by TEM, and understanding of the conventional EM image can be applied. Furthermore, the liver has a hepatic lobule unit, with blood flowing from the interlobular tissue to the central vein, and cell functionality (e.g., ammonia purification) varies depending on its localization. Thus, SEM imaging of the entire lobule structure is expected to be used in the future to analyze the combination of morphology and local cell function.

The wide-area SEM technology can therefore provide a big data of ultra-fine morphology images of biological tissues. We call this morphomics analysis 'Morphomics' to comprehensively acquire hyperfine shape information and perform omics analysis.

\subsection{Stereo EM technology}

Steric communication of a large number of cells allows living tissues to exhibit and maintain their function. In addition, stereoscopic EM techniques are being developed that can analyze biological tissue three dimensionally at nano-level resolution. In particular, stereomicroscopic analysis is actively being conducted as part of the Connectome project to clarify the network of neurons in the whole brain \cite{Mikula2015}.

One of the methodologies for achieving stereo EM is the array tomography (AT) method. In the AT method, ultra-thin sections are continuously cut out and collected from resin-embedded biological tissue samples. Then, after staining, the same site of each section is sequentially observed by SEM, and a continuous tomographic image is obtained from each section, with final three-dimensional reconstruction performed using computer software. One advantage of the AT method is that it can be applied to relatively large tissue samples. Furthermore, since the tissue section remains after observation, re-observation of the same area or another area can be performed. However, at present, it is difficult to manually prepare continuous ultra-thin sections of hundreds to thousands of samples. Further research, with the development and popularization of an advanced microtome, as well as a device for automatically collecting tissue sections, are desired.

Finally, as another method, a block surface observation method using a focused ion beam SEM and a serial block-face SEM, in which a microtome is incorporated in the SEM \cite{Denk2004}, can also provide important information. In recent years, using these stereomicroscopy techniques, Ichimura et al. reported in detail the three-dimensional structure of the complex foot process of podocytes surrounding the periphery of the kidney glomerulus \cite{Ichimura2017}. In studies using stereo EM, the three-dimensional structures of normal tissue structures and associated changes during pathological conditions are examined at the electron microscope level. Thus, it is expected that these techniques will lead to a deeper understanding that could not otherwise be obtained using conventional EM techniques.

\subsection{Functional EM}

Understanding of the living body can be further deepened if tissue function can be estimated at the nano-level by EM. For this purpose, a technique capable of (1) merging gene expression patterns and EM images and (2) merging elemental information and EM images is desired. Although there are methods, such as immuno-EM and FISH in EM analysis, there is no general methodology yet; thus, the development of this technology is under way. In addition, in recent years, the development of cultured cells with the reporter gene APEX2 introduced that can be detected as an EM guide has been undertaken \cite{Martell2017}.

On the other hand, energy-dispersive X-ray (EDX) analysis is a methodology for acquiring elemental information patterns on a sample surface. By applying EDX to EM analysis, the functional state of cells from the array patterns of multiple contained elements can be inferred. Recently, Scotuzzi et al. has shown that EDX analysis of the rat pancreas can distinguish cytoplasmic mitochondria, granules, etc., through elemental fingerprinting \cite{Scotuzzi2017}. Therefore, in the future, the analysis of disease tissues may also proceed by applying such a functional electron microscope.

\subsection{Application of the latest EM technology to histopathology}

Electron microscopic pathology in Japan is routinely performed on kidney biopsy tissue, endocrine tumors, soft tissue malignancies, lipid storage diseases such as Gaucher disease, polysaccharide storage diseases, and myocardial biopsy. However, microscopic examination is primarily performed using light microscopy, and electron microscopic diagnosis is regarded as auxiliary at the present time. 

In most cases, several problems with conventional EM exist: (1) the region that can be observed with EM is extremely narrow; (2) the requesting doctor cannot observe the whole of the EM views and findings; and (3) EM has no quantitative or objective index. These issues are expected to be solved by the advent of innovative technologies such as the wide-area electron microscope introduced in this review, which is (1) able to observe biological tissues of several millimeters square or more, (2) the requesting doctor can confirm all image data through the web browser by utilizing the wide-area EM image viewer, and (3) rapid advances in information processing technology make it possible to handle image big data quantitatively.

We believe that in the future new pathologies can be explored by utilizing wide-area EM, stereo EM, and functional EM for pathological diagnosis. In addition, we consider that the clinical and diagnostic value of EM diagnosis will increase dramatically. Using kidney disease as an example, many kidney biopsies are widely performed at hospitals throughout Japan. The sample blocks of biopsy specimens stored in universities and hospitals can be applied to these EM methods. Moreover, by linking with the accumulated clinical data, it is expected that the clinical significance of the biological structures of the lesion site observed by wide-area EM will become clear. 

In recent years, with the development of a multi-beam-type SEM, imaging speed has greatly improved \cite{EBERLE2015microscopy}. In addition, Toyooka et al. are currently developing a technology for wide-area TEM \cite{Toyooka2014}. Furthermore, a large-area observation by TEM using a silicon nitride window chip has been reported \cite{Konyuba2018}. We hope that the application of EM techniques to histopathological analysis will create a new path for the field of pathology.

\section{Information infrastructure for Image Big data analysis}

New imaging technology called wide-area EM is emerging for morphomics analysis, which comprehensively visualizes and quantifies the morphological information of biological tissues (Table 1). In recent years, the acquisition speed of imaging data has dramatically accelerated. Indeed, it is being produced at a level comparable with the omics data obtained with next-generation sequencing. Thus, the era of image big data is already approaching. It has already been reported that single-beam SEM can acquire several tens of gigabytes (GB) in a single day of imaging, whereas multi-beam SEM can acquire hundreds GB or more of image data. There is an urgent need to build an operation basis and analysis method for such image big data.

\subsection{Image data operation base and information integration}

The image big data era requires fundamentally different data management from conventional human manual image information management. Furthermore, as with human genome information, open data and data accessibility are important, thus allowing anyone in the world to access data on the Web. For that purpose, it is necessary to (1) construct a descriptive format of information science that can read image data directly with a computer, and (2) construct a mechanism to distribute the image data as open data on the Web.

Our research group has experimentally released an image database for large-area EM images using next-generation web technology called Semantic Web and Linked Open Data (LOD), which are distributed in the machine-readable web form \cite{Kume2017}. Specifically, We used this based on the data model of the Open Microscopy Environment \cite{Swedlow2007, Swedlow2020}. We have developed metadata that enables the consistent analysis of samples, pretreatment, staining conditions, imaging conditions, imaging methods, and web data management. Part of these results is the development of an image metadata system in cooperation with the large-area EM image web viewer, and the verification of the practicability of the system. In the future, we believe that the application to medical imaging fields, such as MRI and PET/CT, with human tissue samples and pathological tissues is also important.

The biggest advantage of open data conversion is that we can integrate different field data to realize comprehensive knowledge. Data integration connects data on the web with meaningful links, called semantic web, to form a network. In the modern web environment, data linkage and data integration can be easily realized by linking the data of various fields published worldwide. Such a mechanism is also called LOD and has become one of the trends in realizing the open data world. In addition, an international research group led by the University of Dundee in Scotland is working on making public image archives in the medical sciences and life sciences \cite{Williams2017}. It is expected that open data conversion of the image data will become mainstream in the field of biology and medicine.

\subsection{Artificial intelligence technology in image data analysis}

Recently, image data analysis using artificial intelligence technology, such as machine learning, has received a great deal of attention. Among the image data, EM images belong to a class in which data analysis is difficult. In addition, there has been little progress in the analytical methods. However, in the field of neuroscience research, the Human Connectome Project has begun in earnest. Then, deep learning, which uses a multi-layered neural network for a learning device is applied, and automation and quantification such as the segmentation of neural cells can be started. Furthermore, in the field of pathology, it has also been reported that gene mutation patterns are derived from the morphology data of lung cancer tissue obtained by light microscopy. In addition, for the rapid diagnosis of acute neurological diseases such as stroke and hydrocephalus, head CT images can be screened at high speed. Thus, the power of artificial intelligence analysis is truly astonishing.

As automatic analysis technologies using artificial intelligence develop, the amount of data to be analyzed is expected to increase dramatically. It is also expected that various different types of field data will be integrated and analyzed as the LOD is promoted to promote data linkage. By promoting data collaboration, it is also expected that various different types of field data will be integrated and analyzed. Under these circumstances, data-driven research that can discover new findings from big data will accelerate in the future. The pathophysiology of the future may be fighting with image big data, LOD, and machine learning in front of a computer rather than looking at a tissue section through a microscope.

\section{Conclusion}

In pathology and histology, observations using light microscopy remain important. However, rapid advances in various measurement devices and software have led to the emergence of innovative imaging technologies that were not anticipated several years ago. This review outlines new imaging technologies, including wide-area EM, as well as information-based technologies for handling image big data. It is essential to further examine whether these techniques can be considered the second and third pathologic diagnosis methods following light microscope.

Novel imaging technology opens up new pathological boundaries by comprehensively and efficiently observing biological tissues without bias. The Image Big Data era is a challenge as it is yet unexplored. There is no researcher who foresees the goal or the new discovery just yet. It is impervious, harmful, and worthless to divide the shallow wisdom, as if you know the world of the unexplored. I believe that there are various things in living tissues that we have never thought of before. For the time being, the authors hope to enjoy the full-fledged search of pathology as a basic researcher.

\begin{table}
 \caption{Comparison of imaging techniques using electron beams (subject is assumed to be animal samples).}
 \label{table:SpeedOfLight}
 \centering
  \begin{tabular}{|p{4.5cm}|p{4.5cm}|p{4.5cm}|}
   \hline
   & Conventional EM & Wide-area EM \\
   \hline \hline
   Apparatus & Mainly TEM & Mainly SEM \\
   \hline
   Feature & Narrow observation & Wide-range observation is possible by tiling imaging \\
   \hline
   Target sample & 
   Mainly cell observation and mainly applied to ultra-thin sections of tissue (50 to 100 nm) & Mainly the entire organizational structure observation and mainly applied to ultra-thin and thick sections of tissue (several hundred $\mu$m to several mm) \\
   \hline
   Resolution & nm order (Super-resolution imaging is possible) & nm order \\
   \hline
    Imaging range & Less several tens of $\mu$m square & Several tens of $\mu$m to several mm square \\
    \hline
    Long observation & Little difficult Mainly using grit for TEM & Easy Mainly using silicon wafer \\
    \hline
    3D stereo analysis & Difficult & Possible (AT method etc.) \\
    \hline
    Functional analysis & Little difficult & Possible by EDX and metal sensitization \\
   \hline
  \end{tabular}
\end{table}

\bibliographystyle{unsrtnat}
\bibliography{references}  

\end{document}